\documentclass[aps,prb,twocolumn,preprintnumbers,amsmath,amssymb,superscriptaddress]{revtex4-1}

\usepackage{graphicx}
\usepackage{bm}
\usepackage[dvipdfmx,colorlinks=true,citecolor=blue,linkcolor=blue,filecolor=blue,urlcolor=blue]{hyperref}

\begin{document}

\title{Origin of Pressure-induced Superconducting Phase in $\mathrm{K}_{x}\mathrm{Fe}_{2-y}\mathrm{Se}_{2}$ studied by Synchrotron X-ray Diffraction and Spectroscopy}

\author{Yoshiya Yamamoto} 
  \affiliation{Graduate School of Science and Technology, Kwansei Gakuin University, 2-1 Gakuen, Sanda, Hyogo 669-1337, Japan}
\author{Hitoshi Yamaoka}
  \affiliation{RIKEN SPring-8 Center, RIKEN, 1-1-1 Kouto, Mikazuki, Sayo, Hyogo 679-5148, Japan}
\author{Masashi Tanaka}
  \affiliation{MANA, National Institute for Materials Science, 1-2-1 Sengen, Tsukuba, Ibaraki 305-0047, Japan}
\author{Hiroyuki Okazaki}
  \affiliation{MANA, National Institute for Materials Science, 1-2-1 Sengen, Tsukuba, Ibaraki 305-0047, Japan}
  \affiliation{Advanced Institute for Materials Research, Tohoku University, 2-1-1 Katahira, Aoba, Sendai, Miyagi 980-8577, Japan}
\author{Toshinori Ozaki}
  \affiliation{MANA, National Institute for Materials Science, 1-2-1 Sengen, Tsukuba, Ibaraki 305-0047, Japan}
  \affiliation{Graduate School of Science and Technology, Kwansei Gakuin University, 2-1 Gakuen, Sanda, Hyogo 669-1337, Japan}
\author{Yoshihiko Takano}
  \affiliation{MANA, National Institute for Materials Science, 1-2-1 Sengen, Tsukuba, Ibaraki 305-0047, Japan}
\author{Jung-Fu Lin}
  \affiliation{Department of Geological Sciences, The University of Texas at Austin, Austin, Texas 78712, USA}
  \affiliation{Center for High Pressure Science and Technology Advanced Research (HPSTAR), Shanghai 201203, China}
\author{Hidenori Fujita}
  \affiliation{Center for Science and Technology under Extreme Conditions (KYOKUGEN), 
Graduate School of Engineering Science, Osaka University, Toyonaka, Osaka 560-8531, Japan}
\author{Tomoko Kagayama}
  \affiliation{Center for Science and Technology under Extreme Conditions (KYOKUGEN), 
Graduate School of Engineering Science, Osaka University, Toyonaka, Osaka 560-8531, Japan}
\author{Katsuya Shimizu}
  \affiliation{Center for Science and Technology under Extreme Conditions (KYOKUGEN), 
Graduate School of Engineering Science, Osaka University, Toyonaka, Osaka 560-8531, Japan}
\author{Nozomu Hiraoka}
  \affiliation{National Synchrotron Radiation Research Center, Hsinchu 30076, Taiwan}
\author{Hirofumi Ishii}
  \affiliation{National Synchrotron Radiation Research Center, Hsinchu 30076, Taiwan}
\author{Yen-Fa Liao}
  \affiliation{National Synchrotron Radiation Research Center, Hsinchu 30076, Taiwan}
\author{Ku-Ding Tsuei}
  \affiliation{National Synchrotron Radiation Research Center, Hsinchu 30076, Taiwan}
\author{Jun'ichiro Mizuki}
  \affiliation{Graduate School of Science and Technology, Kwansei Gakuin University, 2-1 Gakuen, Sanda, Hyogo 669-1337, Japan}

\date{\today}

%\keywords{Keyword1, Keyword2, Keyword3}

\begin{abstract}
Pressure dependence of the electronic and crystal structures of $\mathrm{K}_{x}\mathrm{Fe}_{2-y}\mathrm{Se}_{2}$, which has pressure-induced two superconducting domes of SC I and SC II, was investigated by x-ray emission spectroscopy and diffraction. 
X-ray diffraction data show that compressibility along the $c$-axis changes around 12 GPa, where a new superconducting phase of SC II appears. 
This suggests a possible tetragonal to collapsed tetragonal phase transition. 
X-ray emission spectroscopy data also shows the change in the electronic structure around 12 GPa. 
These results can be explained by the scenario that the two SC domes under pressure originate from the change of Fermi surface topology.
Present results here show that the nesting condition plays a key role in stabilizing the superconducting state helping to address outstanding fundamental question as to why the SC II appears under pressure.
\end{abstract}

\maketitle

\section*{Introduction}
Since the discovery of high-temperature superconductivity in F-doped LaFeAsO in 2008,\cite{Kamihara08} many iron-based superconductors with different crystal structures have been synthesized and are still hot topics in condensed matter physics.
Most iron-superconductor families have FeAs or FeSe planes as the common layers, which correlate to the superconductivity.
The crystal structure of FeSe is the simplest of these iron-based superconductors with $T_{\mathrm{c}}=8 \mathrm{K}$.\cite{Hsu08}
Moreover, it was recently found that a single FeSe layer on $\mathrm{SrTiO}_{3}$ showed high $T_{\mathrm{c}}$ of 65--100 K.\cite{He13,Ge15}

Intercalation to FeSe layers by alkaline atoms also raised  $T_{\mathrm{c}}$ to 30--46 K in bulk iron-based superconductors of $A_{x}\mathrm{Fe}_{2-y}\mathrm{Se}_{2}$ ($A=\mathrm{K}, \mathrm{Rb}, \mathrm{Cs}$).\cite{Guo10,Liu11,Ying12,Dagotto13}
Therefore, in these systems electron-doping to the FeSe layer may play an important role in superconductivity. 
The electron-doping causes a Fe-deficiency of the FeSe layer to keep the charge valance, and this system is called 122* phase. 
These new iron-defected systems (122* family) have attracted many interests because of the following several unique features, which are very different from other iron-based superconductors.\cite{Dagotto13} 
(i) This system shows intrinsic phase separation.\cite{Ricci15}
It consists of 122-type superconductor $\mathrm{K}\mathrm{Fe}_{2}\mathrm{Se}_{2}$ and 245-type AFM insulator $\mathrm{K}_{2}\mathrm{Fe}_{4}\mathrm{Se}_{5}$ with $\sqrt{5}\times\sqrt{5}$ vacancy order which disappears around 10GPa.\cite{FChen11,Ding13,Li12,Bendele14,Saini14,Guo12}
(ii) They have an unprecedented high N{\'e}el temperature of 559 K and large magnetic moment of $\sim$3.3$\mu_{\mathrm{B}}$.~\cite{Bao11,Ye11} This magnetic moment is the largest among pnictide and chalcogenide iron-superconductors. 
(iii) Unlike the usual iron-based superconductor, there are no hole pockets at Fermi surface which enhances the Fermi surface nesting.\cite{Zhang11} 
(iv) $T_{\mathrm{c}}$ of $A_{x}\mathrm{Fe}_{2-y}\mathrm{Se}_{2}$ gradually drops with pressure, and superconductivity (SC I) disappears around 10 GPa. However, interestingly, further pressure induces a new superconductivity (SC II) suddenly around 11 GPa. The SC II phase shows higher $T_{\mathrm{c}}$ than the SC I phase.\cite{Sun12,Gao14}

Recently, single phase non-superconducting $\mathrm{K}_{2}\mathrm{Fe}_{4}\mathrm{Se}_{5}$ was synthesized, and the pressure-temperature phase diagram was revealed.\cite{Gao14}
By comparing the $\mathrm{K}_{2}\mathrm{Fe}_{4}\mathrm{Se}_{5}$ and $\mathrm{K}_{x}\mathrm{Fe}_{2-y}\mathrm{Se}_{2}$ phase diagrams, the phase separation in the SC II region was suggested, and the superconducting phase attributed to the 122 phase.
Therefore, this means that superconducting phase with $\mathrm{K}\mathrm{Fe}_{2}\mathrm{Se}_{2}$ and non-superconducting phase with $\mathrm{K}_{2}\mathrm{Fe}_{4}\mathrm{Se}_{5}$ co-exist in the SC II phase.

A theoretical study of the SC I and SC II phases in the 122* system suggested that superconducting symmetry is $d$-wave without $\Gamma$-point hole pocket at SC I and $s_{\pm}$-pairing at SC II.\cite{Das13} 
In these systems, however, since no experimental study of not only the electronic structure, but also the crystal structure under pressure has been reported so far, the issue of the appearance of SC II dome under pressure remains unclear. 

In this paper we report a systematic study of $\mathrm{K}_{x}\mathrm{Fe}_{2-y}\mathrm{Se}_{2}$ with x-ray diffraction (XRD) and x-ray emission spectroscopy (XES) under pressure. 
The purpose of this work is to reveal both the crystal and electronic structures of $\mathrm{K}_{x}\mathrm{Fe}_{2-y}\mathrm{As}_{2}$ under pressure to clarify the existence of the two superconducting domes.  
XES technique has made it possible to probe local magnetic moment under pressure by detecting Fe $K\beta$ emission spectra for iron-based superconductor.~\cite{Tsutsumi59,Vanko06,JMChen11,Gretarsson11,Gretarsson13} 
We also performed the bulk sensitive x-ray absorption (XAS) measurements with partial fluorescence (PFY) mode at the Fe $K$ absorption edge.~\cite{Hamalainen91}  
We have used the PFY-XAS method where a decay process with shorter life time is selected, resulting spectra are narrower, and making fine electronic structure near the absorption edge better visible.\cite{Hamalainen91,Dallera02,Yamaoka14} 
Our results show the change in the $c$-axis compressibility around boundary pressure of the SC I and SC II phases, suggesting a crystal structure change at this pressure, probably a tetragonal (T) to collapsed tetragonal (cT) transition. 
The Fe $K\beta$ XES also shows a pressure-induced change in the electronic structure at the transition pressure. 

\section*{Results}
\subsection*{$P$-$T$ phase diagram}
\begin{figure}
  \includegraphics{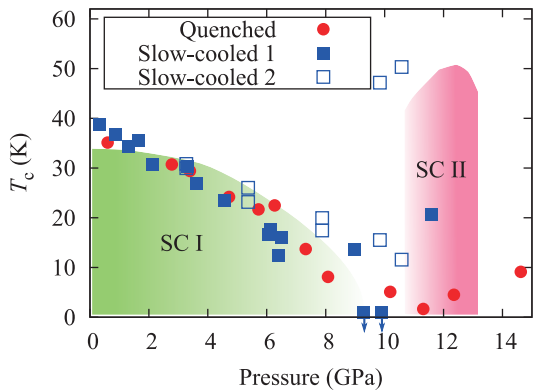}
  \caption{(Color online) A $P$-$T$ phase diagram of $\mathrm{K}_{x}\mathrm{Fe}_{2-y}\mathrm{Se}_{2}$.\cite{Fujita15} Red circles and blue squares indicate quenched sample and slow-cooled sample, respectively. Colouring region is based on the data taken from the reference.\cite{Sun12}}
  \label{fig:Tc}
\end{figure}
We prepared two kinds of $\mathrm{K}_{x}\mathrm{Fe}_{2-y}\mathrm{Se}_{2}$ single crystals: a sample quenched at 550 $^{\circ}$C (quenched sample) and one cooled slowly (slow-cooled sample). 
A $P$-$T$ phase diagram of the quenched and slow-cooled samples is shown in Fig.~\ref{fig:Tc}.
$T_{\mathrm{c}}$ was determined from the onset temperature of the electrical resistivity measurements.
Both samples show the $T_{\mathrm{c}}$ decreases with pressure monotonically in the SC I phase. This behavior agrees well with the reports published.\cite{Sun12,Gao14}
However, the maximum $T_{\mathrm{c}}$ of SC II phase depends on the samples. $T_{\mathrm{c}}$ of the quenched and slow-cooled samples are $\sim$ 5 K and $\sim$ 20 K at the SC II phase, respectively,~\cite{Fujita15} while $T_{\mathrm{c}}$ of SC II was $\sim$ 50 K in the reports published.\cite{Sun12,Gao14}
These results suggest that the $T_{\mathrm{c}}$ of SC II depends strongly on the sample preparation.
Actually, island- and mesh-shape morphology were observed in the back-scattered electron (BSE) image in the slow-cooled and the quenched samples, respectively.\cite{Tanaka15}
These morphologies were caused by the difference of iron concentration.\cite{Tanaka15}

\subsection*{X-ray diffraction}
\begin{figure*}
  \includegraphics{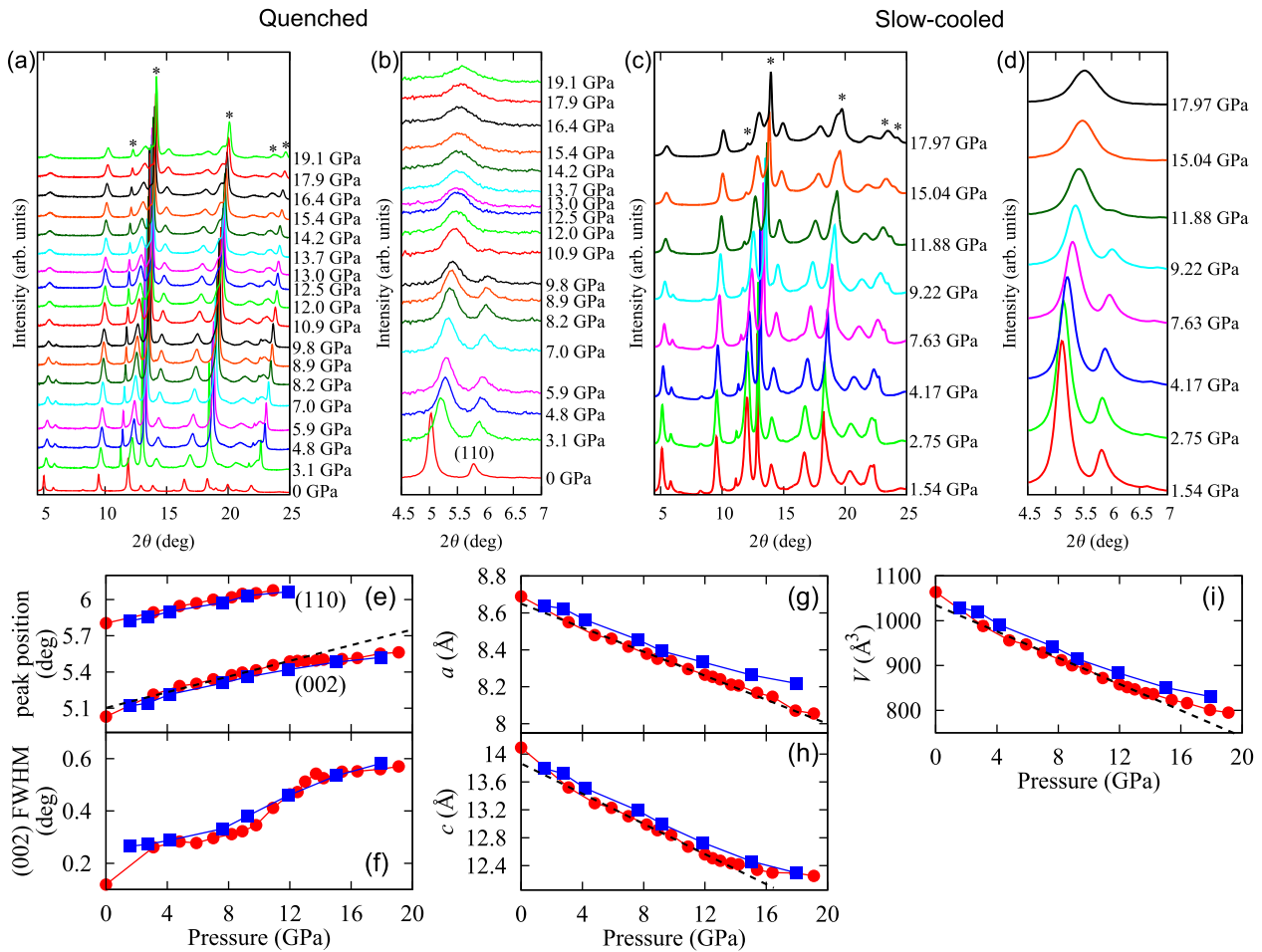}
  \caption{(Color online) XRD pattern of (a) the quenched sample and (c) the slow-cooled sample. (b) and (d) Enlarged views of (a) and (c), respectively. Asterisk mark means reflection of NaCl used as the pressure medium of the diamond anvil cell. In the both quenched and slow-cooled samples, the (110) superstructure reflection disappear around 12 GPa. (e-i) Pressure evolution of the peak properties and the structure parameters of the quenched (red circle) and slow-cooled (blue square) samples. (e) Peak position of (002) and (110). (f) Full width at half maximum of the (002) peak. (g) Lattice constant along the $a$-axis. (h) Lattice constant along the $c$-axis. (i) Volume. Linear dashed-lines are guides for the eye.}
  \label{fig:XRD}
\end{figure*}
We measured x-ray diffraction patterns under pressure up to 19.1 GPa for the quenched sample and 18.0 GPa for the slow-cooled sample at room temperature as shown in Fig.~\ref{fig:XRD}. 
Both samples consist of a $I4/m$ symmetry of the 245 phase and a $I4/mmm$ symmetry of the 122 phase at ambient pressure. 
Fe vacancy order-disorder transition was reported in the non-superconducting 245 phase at SC II, and crystal symmetry after the transition becomes $I4/mmm$ which is the same as the superconducting phase.\cite{Guo12, Bendele13}
Figures~\ref{fig:XRD}(a) and \ref{fig:XRD}(c) show the XRD pattern of the quenched and slow-cooled samples, and the enlarge views are shown in Figs.~\ref{fig:XRD}(b) and \ref{fig:XRD}(d).
Intensity of the superstructure peak (110) attributed to the Fe vacancy order disappears around 12 GPa, indicating a clear structural phase transition from $I4/m$ to $I4/mmm$ symmetry at 245 phase.  
The same feature has been observed previously.\cite{Guo12, Bendele13}
Seemingly, the above structural transition pressure of 12 GPa coincides with the appearance of the SC II phase as seen in Fig.~\ref{fig:Tc}. 

Although a Rietveld refinement was not performed because of the restriction of the observed $\bm{Q}$ range, 
we performed peak fits by using the several peaks with the Voigt functions in order to derive the lattice constants. 
Figure~\ref{fig:XRD}(e) indicates (002) and (110) peak position vs pressure.
Trend of the pressure evolution of (002) peak position changes around 12 GPa.
This system consists of the 122 and 245 phases and thus only the average lattice constant of two phases could be analyzed. 
Here, we assumed $I4/m$ symmetry at all pressures because $I4/mmm$ symmetry can express $I4/m$ symmetry.
Figures~\ref{fig:XRD}(e) and \ref{fig:XRD}(f) show pressure evolution of the lattice constants.
Pressure evolution of the $a$-axis shows a monotonic decrease, while that of the $c$-axis changes the slope around 12 GPa.
Thus the compressibility along the $c$-axis becomes lower above 12 GPa.
This means that the bond along the $c$-axis at the SC II phase is stronger than that at the SC I phase.
This suggests a crystal structure change at 12 GPa, probably T $\to$ cT structural phase transition analogous to $\mathrm{Eu}\mathrm{Fe}_{2}\mathrm{As}_{2}$.\cite{Yu14}

\subsection*{Pressure induced change in the $K\beta$ emission spectra}
\begin{figure*}
  \includegraphics{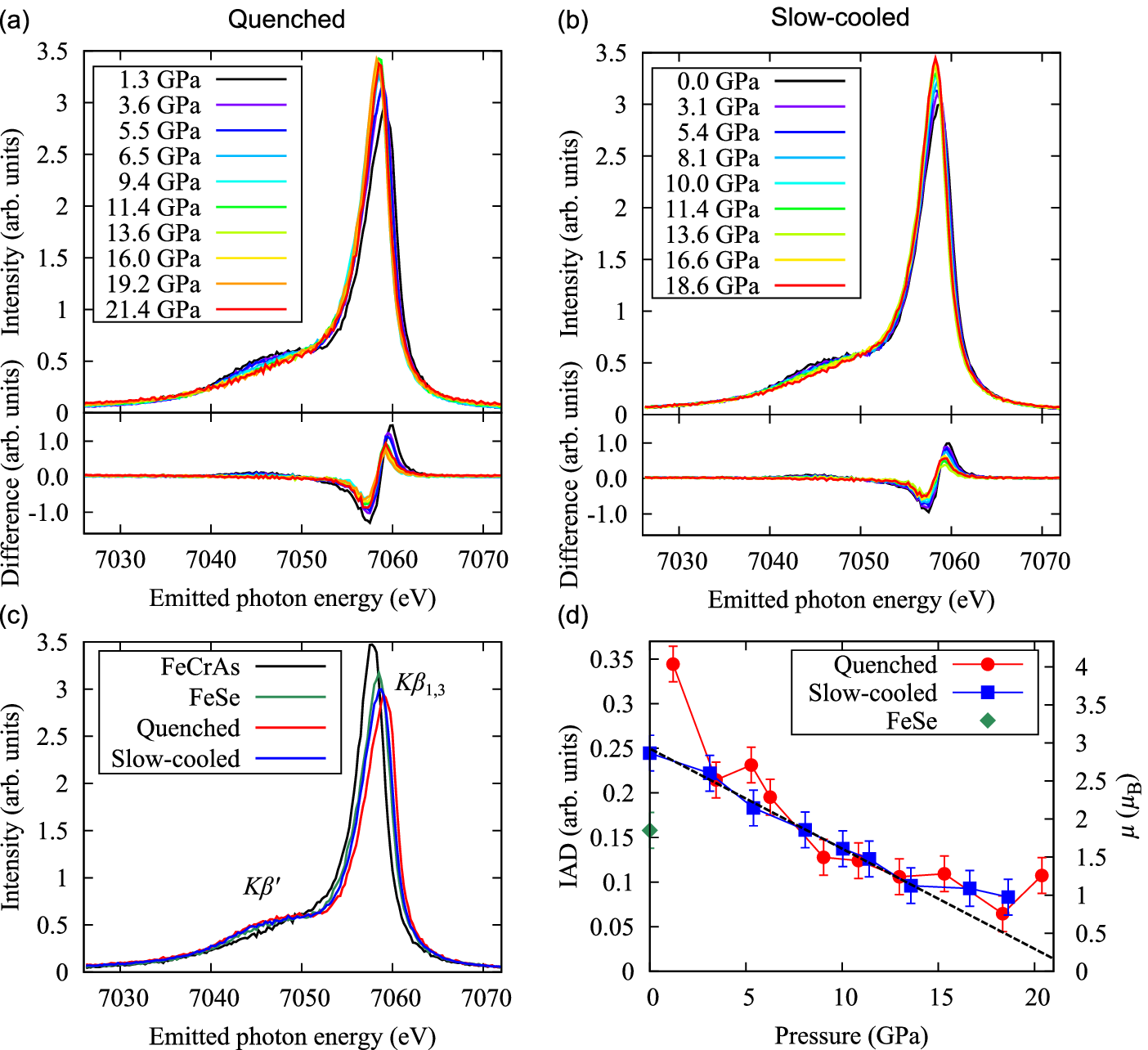}
  \caption{(Color online) Pressure dependence of $K\beta$ emission spectra of the (a) quenched and (b) slow-cooled samples. (c) $K\beta$ spectra of FeCrAs, FeSe, the quenched sample, and the slow-cooled sample. (d) Pressure dependence of amplitude of magnetic moment per Fe estimated with the IAD values of the $K\beta$ spectra. A linear dashed-line is a guide for the eye. }
  \label{fig:kb+IAD}
\end{figure*}
Figures~\ref{fig:kb+IAD}(a) and \ref{fig:kb+IAD}(b) show pressure evolution of $K\beta$ emission spectra of the quenched and slow-cooled samples, respectively.
A $K\beta$ spectrum consists of a main peak of $K\beta_{1,3}$ and a satellite peak of $K\beta'$, which correspond to low-spin and high-spin states, respectively.\cite{Tsutsumi59}
In Fig.~\ref{fig:kb+IAD}, pressure evolution of $K\beta$ spectrum shows a shift from the high-spin to the low-spin state with pressure.

Figure~\ref{fig:kb+IAD}(c) shows a comparison among the $K\beta$ spectra of the quenched sample, the slow-cooled sample, FeCrAs (0$\mu_{\mathrm{B}}$), and FeSe (2$\mu_{\mathrm{B}}$).
As seen in Fig.~\ref{fig:kb+IAD}(c), comparison of $K\beta$ spectra between $\mathrm{K}_{x}\mathrm{Fe}_{2-y}\mathrm{Se}_{2}$ and FeCrAs concludes that $\mathrm{K}_{x}\mathrm{Fe}_{2-y}\mathrm{Se}_{2}$ is in the higher-spin state because of larger $K\beta'$ intensity.
The local moment of Fe can be extracted by the the integrated absolute difference (IAD) analysis of the Fe $K\beta$ emission spectra to a reference spectrum.\cite{Vanko06,Gretarsson11}
It is known that the IAD values are proportional to the local magnetic moments.\cite{Gretarsson11} 

Figure~\ref{fig:kb+IAD}(d) shows the local magnetic moment estimated by the IAD analysis of the $K\beta$ spectra in Figs.~\ref{fig:kb+IAD}(a) and \ref{fig:kb+IAD}(b).
The local magnetic moment decreases from $\sim 3\mu_{\mathrm{B}}$ at ambient pressure to $\sim 1\mu_{\mathrm{B}}$ at the SC II phase with pressure. 
Two samples show roughly the same trend under pressure.
Especially the pressure evolution of the local magnetic moment of slow-cooled sample changes the slope at 12 GPa.
This coincides with the change in the compressibility along the $c$-axis shown in Fig.~\ref{fig:XRD}(h).

\subsection*{Pressure induced change in the PFY-XAS spectra}
\begin{figure*}
  \includegraphics{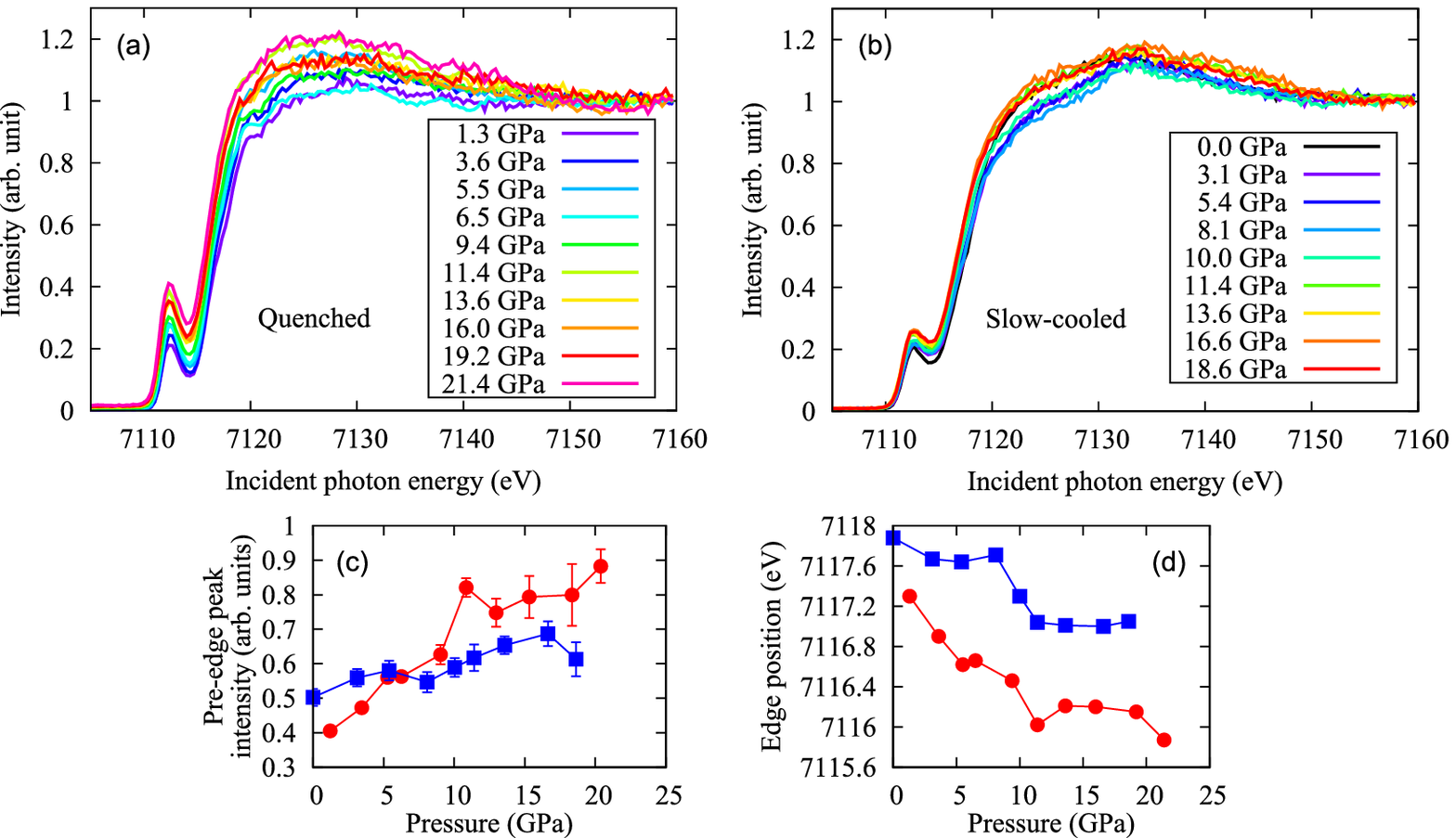}
  \caption{(Color online) Pressure evolution of the PFY-XAS spectra of (a) the quenched sample and (b) the slow-cooled sample. In both the quenched and slow-cooled samples, the pre-edge peak intensity increase with pressure and edge position move towered to low energy. Pressure evolution of (c) the pre-edge peak intensity and (d) the edge position. Red circle and blue square indicate the quenched and slow-cooled samples, respectively.}
  \label{fig:PFY-XAS}
\end{figure*}
Figures~\ref{fig:PFY-XAS}(a) and \ref{fig:PFY-XAS}(b) show a pressure evolution of the PFY-XAS spectra setting the emitted photon energy to $K\beta_{1,3}$ peak of the quenched and slow-cooled samples, respectively. 
The intensity is normalized to that at 7160 eV. 
The PFY-XAS spectra show large pre-edge peaks.
The pre-edge and the main edge peaks correspond to $1s \to 3d$ quadrupole and  $1s \to 4p$ dipole transitions, respectively.
The strong pre-edge peak intensity includes the information of the hybridization between the Fe $3d$ and Se $4p$ orbitals.\cite{Groot09}
The edge position of the PFY-XAS spectra shifts toward lower energy with pressure in both samples, indicating the decrease of the Fe valence. 
The system includes $\mathrm{Fe}^{3+}$ and $\mathrm{Fe}^{2+}$\cite{Simonelli12} and thus the above result indicates a change in the Fe valence from $\mathrm{Fe}^{3+} \to \mathrm{Fe}^{2+}$.
The decrease of the Fe valence with pressure may be due to the electron supply from K to FeSe layer caused by the shrink along the $c$-axis.
Figures~\ref{fig:PFY-XAS}(c) shows that the pre-edge peak intensity of the PFY-XAS spectra increases with pressure.
Another point we would like to emphasize is that the intensity around 7125 eV changes at 12GPa in the slow-cooled sample, although it is not clear in the quenched one (Fig.~\ref{fig:PFY-XAS}(a) and (b)).
This pressure also coincides with the pressure where the compressibility of the $c$-axis changes.

\section*{Discussion}
The XRD and XES studies under pressure have been performed for the 122* system, which have pressure-induced two superconducting domes. 
The XRD results show that the compressibility along the $c$-axis changes at 12 GPa and the superlattice diffraction disappears at the same pressure.
Pressure dependence of the lattice constant along the $c$-axis and the volume becomes gentle at the SC II phase.
The same $c$-axis evolution has been observed in $A\mathrm{Fe}_{2}\mathrm{As}_{2}$ where $A=$ Ca, Sr, Ba and Eu.\cite{Yu14}
This was interpreted as the T $\to$ cT structural phase transition.
Therefore, it is reasonable to expect that $\mathrm{K}_{x}\mathrm{Fe}_{2-y}\mathrm{Se}_{2}$ system, which has the same crystal structure, also shows the T $\to$ cT transition.

The change in the crystal structure affects the magnetic property. 
Actually, the $K\beta$ XES results indicate that the trend of the pressure evolution of the magnetic moment and the electronic state shown in Fig.~\ref{fig:kb+IAD}(d) changes also at 12 GPa, which seems to correlate to the T $\to$ cT transition.
The average local magnetic moment changes from $\sim 3\mu_{\mathrm{B}}$ at ambient pressure to $\sim 1\mu_{\mathrm{B}}$ at the SC II phase with pressure.
The change in the magnetic moment at 12 GPa is not large in $\mathrm{K}_{x}\mathrm{Fe}_{2-y}\mathrm{Se}_{2}$ system, probably because the collapse along the $c$-axis at 12 GPa is small.

The PFY-XAS spectra show that the Fe valence decreases with pressure, which may correspond to the increase of the carrier density at the SC II phase due to the supply of the electrons from K to the FeSe layer caused by the shrink along the $c$-axis.
The pre-edge peak intensity in the PFY-XAS spectra increases with pressure, indicating the increase of the hybridization between Fe $3d$ and Se $4p$ and also the density of states (DOS) near the Fermi surface.
The pressure-induced change in the pre-edge peak intensity also correlates to the shift from high-spin to low-spin states.

In the 122 system the superconductivity emerged suddenly at the cT phase when the T $\to$ cT structural phase transition occurred.\cite{Ying15}
The phase diagram of the 122 system is similar to that of the 122* system.\cite{Nakajima15}
The DFT calculations showed the change in the electronic structure between the T phase and cT phase.\cite{Guterding15}
In $\mathrm{KFe}_{2}\mathrm{As}_{2}$, the T $\to$ cT transition changed the superconductivity symmetry from $d$-wave to $s$-wave. 
This is a Lifshitz transition which is known to change the Fermi surface drastically from the electronic state with only hole pocket to that with electron and hole pockets. 
Other calculations of $\mathrm{K}_{x}\mathrm{Fe}_{2-y}\mathrm{Se}_{2}$ also showed the $d$-wave in the SC I and $s$-wave in the SC II phase.\cite{Das13}
Therefore, together with these theoretical calculations we conclude that there is the T $\to$ cT transition in $\mathrm{K}_{x}\mathrm{Fe}_{2-y}\mathrm{Se}_{2}$ under pressure and thus the high-$T_{\mathrm{c}}$ at the SC II phase could be explained by the strong Fermi surface nesting.

\section*{Methods}
\subsection*{Sample preparation and characterizations}
We prepared two kinds of $\mathrm{K}_{x}\mathrm{Fe}_{2-y}\mathrm{Se}_{2}$ single crystals.\cite{Ozaki13,Tanaka15} 
Single crystals were grown by a simple one-step synthesis.
Fe (99.9\%), $\mathrm{K}_{2}\mathrm{Se}$ (99\%) powders and Se (99.999\%) grains were put into an alumina crucible and sealed into an evacuated quartz tube. 
The mixture was slowly heated up to 900 $^{\circ}$C and held for 3 hours. 
The melting mixture was, then, cooled down to room temperature slowly (slow-cooled sample) and quenched at 550 $^{\circ}$C (quenched sample).
Back-scattered electrons (BSE) images were obtained to observe micro-structure. 
Island- and mesh-shape structure were shown in the slow-cooled and quenched samples, and the chemical composition determined by using energy dispersive x-ray spectrometry (EDX) were
$\mathrm{K}_{0.40}\mathrm{Fe}_{1.95}\mathrm{Se}_{2}$ and $\mathrm{K}_{0.63}\mathrm{Fe}_{1.71}\mathrm{Se}_{2}$, respectively.\cite{Tanaka15}
The area ratios between the superconducting region and non-superconducting region is $\sim$10-13\% in the slow-cooled sample and $\sim$30-35\% in the quenched sample.
$T_{\mathrm{c}}$ of the present samples under pressure were measured at Osaka University.\cite{Fujita15}

\subsection*{XRD, XES, and PFY-XAS measurements under pressure}
We performed XRD, XES, and PFY-XAS experiments for the slow-cooled and quenched samples. 
For XRD, XES, and PFY-XAS measurement, these samples with NaCl as the pressure medium were loaded into a sample chamber of the gasket in the glove box of pure Ar atmosphere because these samples are chemically unstable in the air.
Pressure was monitored by ruby fluorescence method.\cite{Mao76}

Pressure dependence of the XRD patterns were measured at SPring-8 BL12B2 using a 3-pin plate diamond anvil cell (DAC, Almax Industries) with a CCD detection system at room temperature. 
We took an arrangement of both incoming and outgoing x-ray beams passed through the diamonds with incident photon energy of 20 keV.
NaCl was loaded as the pressure medium and well-mixed with the sample because of reduction of preferred orientation of the sample.
2D image of CCD was integrated by using FIT2D program.\cite{Hammersley96}

The PFY-XAS and XES measurements were performed at the Taiwan beam line BL12XU at SPring-8.
The undulator beam was monochromatized by a cryogenically-cooled double crystal Si(111) monochromator.
A Johann-type spectrometer equipped with a spherically bent Si(531) analyzer crystal (radius of $\sim$1 m) and a Si solid state detector (Amptech) were used to analyze the Fe emission of the $3p \to 1s$ de-excitation at the Fe $K$ absorption edge.
At the emitted photon energy of 7.6 keV the overall energy resolution was estimated to be 0.9 eV.
The intensities of the measured spectra were normalized using the incident beam that was monitored just before the sample.

For the high-pressure XES experiments the x-ray beam was focused to 20-30 (horizontal) $\times$ 30-40 (vertical) $\mu\mathrm{m}^{2}$ at the sample position using a toroidal and a Kirkpatrick-Baez mirror.
High-pressure conditions were achieved at room temperature using a diamond anvil cell coupled with a gas-membrane.
A Be-gasket with 3 mm in diameter and approximately 100 $\mu\mathrm{m}$ thick was pre-indented to approximately 35-40 $\mu\mathrm{m}$ thickness around the center.
The diameter of the sample chamber in the gasket was approximately 100 $\mu\mathrm{m}$ and the diamond anvil culet size was 300 $\mu\mathrm{m}$.
We used the Be gasket in-plane geometry with a scattering angle of 90$^{\circ}$, where both incoming and outgoing x-ray beams passed through the Be gasket.
Be was used due to its higher transmittance to x-rays in comparison to other high-$Z$ materials.

\subsection*{IAD analyses}
The IAD analysis is performed in the following way: (i) match the center of mass between the sample and reference spectra, (ii) take the difference between them, and (iii) integrate the absolute value of the difference. 
The intensity is normalized by the area of the $K\beta$ spectrum.

%\bibliography{Paper_KxFe2-ySe2}

\begin{thebibliography}{10}
\expandafter\ifx\csname url\endcsname\relax
  \def\url#1{\texttt{#1}}\fi
\expandafter\ifx\csname urlprefix\endcsname\relax\def\urlprefix{URL }\fi
\providecommand{\bibinfo}[2]{#2}
\providecommand{\eprint}[2][]{\url{#2}}

\bibitem{Kamihara08}
\bibinfo{author}{Kamihara, Y.}, \bibinfo{author}{Watanabe, T.},
  \bibinfo{author}{Hirano, M.} \& \bibinfo{author}{Hosono, H.}
\newblock \bibinfo{title}{Iron-based layered superconductor
  $\mathrm{La}[\mathrm{O}_{1-x}\mathrm{F}_{x}]\mathrm{FeAs}$ ($x=0.05$-$0.12$)
  with ${T}_\mathrm{c} = 26\ \mathrm{K}$}.
\newblock \emph{\bibinfo{journal}{J. Am. Chem. Soc.}}
  \textbf{\bibinfo{volume}{130}}, \bibinfo{pages}{3296} (\bibinfo{year}{2008}).
\newblock \urlprefix\url{http://dx.doi.org/10.1021/ja800073m}.

\bibitem{Hsu08}
\bibinfo{author}{Hsu, F.-C.} \emph{et~al.}
\newblock \bibinfo{title}{Superconductivity in the {PbO}-type structure
  $\alpha$-{FeSe}}.
\newblock \emph{\bibinfo{journal}{Proc. Nat. Acad. Sci. (USA)}}
  \textbf{\bibinfo{volume}{105}}, \bibinfo{pages}{14262}
  (\bibinfo{year}{2008}).
\newblock \urlprefix\url{http://dx.doi.org/10.1073/pnas.0807325105}.

\bibitem{He13}
\bibinfo{author}{He, S.} \emph{et~al.}
\newblock \bibinfo{title}{Phase diagram and electronic indication of
  high-temperature superconductivity at 65 {K} in single-layer {FeSe} films}.
\newblock \emph{\bibinfo{journal}{Nat. Mater.}} \textbf{\bibinfo{volume}{12}},
  \bibinfo{pages}{605} (\bibinfo{year}{2013}).
\newblock \urlprefix\url{http://dx.doi.org/10.1038/nmat3648}.

\bibitem{Ge15}
\bibinfo{author}{Ge, J.-F.} \emph{et~al.}
\newblock \bibinfo{title}{Superconductivity above 100 {K} in single-layer
  {FeSe} films on doped $\mathrm{SrTiO}_{3}$}.
\newblock \emph{\bibinfo{journal}{Nat. Mater.}} \textbf{\bibinfo{volume}{14}},
  \bibinfo{pages}{285} (\bibinfo{year}{2015}).
\newblock \urlprefix\url{http://dx.doi.org/10.1038/nmat4153}.

\bibitem{Guo10}
\bibinfo{author}{Guo, J.} \emph{et~al.}
\newblock \bibinfo{title}{Superconductivity in the iron selenide
  $\mathrm{K}_{x}\mathrm{Fe}_{2}\mathrm{As}_{2}$
  $(0\ensuremath{\le}x\ensuremath{\le}1.0)$}.
\newblock \emph{\bibinfo{journal}{Phys. Rev. B}} \textbf{\bibinfo{volume}{82}},
  \bibinfo{pages}{180520} (\bibinfo{year}{2010}).
\newblock \urlprefix\url{http://link.aps.org/doi/10.1103/PhysRevB.82.180520}.

\bibitem{Liu11}
\bibinfo{author}{Liu, R.~H.} \emph{et~al.}
\newblock \bibinfo{title}{Coexistence of superconductivity and
  antiferromagnetism in single crystals
  ${A}_{0.8}\mathrm{Fe}_{2-y}\mathrm{As}_{2}$ (${A}=\mathrm{K}, \mathrm{Rb},
  \mathrm{Cs}$, $\mathrm{Tl}$/$\mathrm{K}$ and $\mathrm{Tl}$/$\mathrm{Rb}$):
  Evidence from magnetization and resistivity}.
\newblock \emph{\bibinfo{journal}{EPL}} \textbf{\bibinfo{volume}{94}},
  \bibinfo{pages}{27008} (\bibinfo{year}{2011}).
\newblock \urlprefix\url{http://stacks.iop.org/0295-5075/94/i=2/a=27008}.

\bibitem{Ying12}
\bibinfo{author}{Ying, T.~P.} \emph{et~al.}
\newblock \bibinfo{title}{Observation of superconductivity at 30-46
  $\mathrm{K}$ in ${A}_{x}\mathrm{Fe}_{2}\mathrm{Se}_{2}$ (${A} = \mathrm{Li}$,
  $\mathrm{Na}$, $\mathrm{Ba}$, $\mathrm{Sr}$, $\mathrm{Ca}$, $\mathrm{Yb}$,
  and $\mathrm{Eu}$)}.
\newblock \emph{\bibinfo{journal}{Sci. Rep.}} \textbf{\bibinfo{volume}{2}},
  \bibinfo{pages}{426} (\bibinfo{year}{2012}).
\newblock \urlprefix\url{http://dx.doi.org/10.1038/srep00426}.

\bibitem{Dagotto13}
\bibinfo{author}{Dagotto, E.}
\newblock \bibinfo{title}{The unexpected properties of alkali metal iron
  selenide superconductors}.
\newblock \emph{\bibinfo{journal}{Rev. Mod. Phys.}}
  \textbf{\bibinfo{volume}{85}}, \bibinfo{pages}{849} (\bibinfo{year}{2013}).
\newblock \urlprefix\url{http://link.aps.org/doi/10.1103/RevModPhys.85.849}.

\bibitem{Ricci15}
\bibinfo{author}{Ricci, A.} \emph{et~al.}
\newblock \bibinfo{title}{Direct observation of nanoscale interface phase in
  the superconducting chalcogenide
  $\mathrm{K}_{x}\mathrm{Fe}_{2-y}\mathrm{Se}_{2}$ with intrinsic phase
  separation}.
\newblock \emph{\bibinfo{journal}{Phys. Rev. B}} \textbf{\bibinfo{volume}{91}},
  \bibinfo{pages}{020503} (\bibinfo{year}{2015}).
\newblock \urlprefix\url{http://link.aps.org/doi/10.1103/PhysRevB.91.020503}.

\bibitem{FChen11}
\bibinfo{author}{Chen, F.} \emph{et~al.}
\newblock \bibinfo{title}{Electronic identification of the parental phases and
  mesoscopic phase separation of
  $\mathrm{K}_{x}\mathrm{Fe}_{2-y}\mathrm{Se}_{2}$ superconductors}.
\newblock \emph{\bibinfo{journal}{Phys. Rev. X}} \textbf{\bibinfo{volume}{1}},
  \bibinfo{pages}{021020} (\bibinfo{year}{2011}).
\newblock \urlprefix\url{http://link.aps.org/doi/10.1103/PhysRevX.1.021020}.

\bibitem{Ding13}
\bibinfo{author}{Ding, X.} \emph{et~al.}
\newblock \bibinfo{title}{Influence of microstructure on superconductivity in
  $\mathrm{K}\mathrm{Fe}_{2-y}\mathrm{Se}_{2}$ and evidence for a new parent
  phase $\mathrm{K}_{2}\mathrm{Fe}_{7}\mathrm{Se}_{8}$}.
\newblock \emph{\bibinfo{journal}{Nat. Commun.}} \textbf{\bibinfo{volume}{4}},
  \bibinfo{pages}{1897} (\bibinfo{year}{2013}).
\newblock \urlprefix\url{http://dx.doi.org/10.1038/ncomms2913}.

\bibitem{Li12}
\bibinfo{author}{Li, W.} \emph{et~al.}
\newblock \bibinfo{title}{Phase separation and magnetic order in {K}-doped iron
  selenide superconductor}.
\newblock \emph{\bibinfo{journal}{Nat. Phys.}} \textbf{\bibinfo{volume}{8}},
  \bibinfo{pages}{126} (\bibinfo{year}{2012}).
\newblock
  \urlprefix\url{http://www.nature.com/nphys/journal/v8/n2/abs/nphys2155.html}.

\bibitem{Bendele14}
\bibinfo{author}{Bendele, M.} \emph{et~al.}
\newblock \bibinfo{title}{Spectromicroscopy of electronic phase separation in
  $\mathrm{K}\mathrm{Fe}_{2-y}\mathrm{Se}_{2}$ superconductor}.
\newblock \emph{\bibinfo{journal}{Sci. Rep.}} \textbf{\bibinfo{volume}{4}},
  \bibinfo{pages}{5592} (\bibinfo{year}{2014}).
\newblock \urlprefix\url{http://dx.doi.org/10.1038/srep05592}.

\bibitem{Saini14}
\bibinfo{author}{Saini, N.~L.} \emph{et~al.}
\newblock \bibinfo{title}{X-ray absorption and photoemission spectroscopy of
  electronic phase separation in
  $\mathrm{K}_{x}\mathrm{Fe}_{2-y}\mathrm{Se}_{2}$}.
\newblock \emph{\bibinfo{journal}{Phys. Rev. B}} \textbf{\bibinfo{volume}{90}},
  \bibinfo{pages}{184510} (\bibinfo{year}{2014}).
\newblock \urlprefix\url{http://link.aps.org/doi/10.1103/PhysRevB.90.184510}.

\bibitem{Guo12}
\bibinfo{author}{Jing, J.} \emph{et~al.}
\newblock \bibinfo{title}{Pressure-driven quantum criticality in iron-selenide
  superconductors}.
\newblock \emph{\bibinfo{journal}{Phys. Rev. Lett.}}
  \textbf{\bibinfo{volume}{108}}, \bibinfo{pages}{197001}
  (\bibinfo{year}{2012}).
\newblock \urlprefix\url{http://dx.doi.org/10.1103/PhysRevLett.108.197001}.

\bibitem{Bao11}
\bibinfo{author}{Bao, W.} \emph{et~al.}
\newblock \bibinfo{title}{A novel large moment antiferromagnetic order in
  $\mathrm{K}_{0.8}\mathrm{Fe}_{1.6}\mathrm{Se}_{2}$ superconductor}.
\newblock \emph{\bibinfo{journal}{Chin. Phys. Lett.}}
  \textbf{\bibinfo{volume}{28}}, \bibinfo{pages}{086104}
  (\bibinfo{year}{2011}).
\newblock \urlprefix\url{http://stacks.iop.org/0256-307X/28/i=8/a=086104}.

\bibitem{Ye11}
\bibinfo{author}{Ye, F.} \emph{et~al.}
\newblock \bibinfo{title}{Common crystalline and magnetic structure of
  superconducting ${A}_{2}\mathrm{Fe}_{4}\mathrm{Se}_{5}$
  (${A}=\mathrm{K},\mathrm{Rb},\mathrm{Cs},\mathrm{Tl}$) single crystals
  measured using neutron diffraction}.
\newblock \emph{\bibinfo{journal}{Phys. Rev. Lett.}}
  \textbf{\bibinfo{volume}{107}}, \bibinfo{pages}{137003}
  (\bibinfo{year}{2011}).
\newblock
  \urlprefix\url{http://link.aps.org/doi/10.1103/PhysRevLett.107.137003}.

\bibitem{Zhang11}
\bibinfo{author}{Zhang, Y.} \emph{et~al.}
\newblock \bibinfo{title}{Nodeless superconducting gap in
  ${A}_{2}\mathrm{Fe}_{2}\mathrm{Se}_{2}$ (${A}=\mathrm{K}$, $\mathrm{Cs}$)
  revealed by angle-resolved photoemission spectroscopy}.
\newblock \emph{\bibinfo{journal}{Nat. Mater.}} \textbf{\bibinfo{volume}{10}},
  \bibinfo{pages}{273} (\bibinfo{year}{2011}).
\newblock \urlprefix\url{http://dx.doi.org/10.1038/nmat2981}.

\bibitem{Sun12}
\bibinfo{author}{Sun, L.} \emph{et~al.}
\newblock \bibinfo{title}{Re-emerging superconductivity at 48 kelvin in iron
  chalcogenides}.
\newblock \emph{\bibinfo{journal}{Nature (London)}}
  \textbf{\bibinfo{volume}{483}}, \bibinfo{pages}{67} (\bibinfo{year}{2012}).
\newblock \urlprefix\url{http://dx.doi.org/10.1038/nature10813}.

\bibitem{Gao14}
\bibinfo{author}{Gao, P.} \emph{et~al.}
\newblock \bibinfo{title}{Role of the 245 phase in alkaline iron selenide
  superconductors revealed by high-pressure studies}.
\newblock \emph{\bibinfo{journal}{Phys. Rev. B}} \textbf{\bibinfo{volume}{89}},
  \bibinfo{pages}{094514} (\bibinfo{year}{2014}).
\newblock \urlprefix\url{http://dx.doi.org/10.1103/PhysRevB.89.094514}.

\bibitem{Das13}
\bibinfo{author}{Das, T.} \& \bibinfo{author}{Balatsk, A.~V.}
\newblock \bibinfo{title}{Origin of pressure induced second superconducting
  dome in ${A}_{y}\mathrm{Fe}_{2-x}\mathrm{Se}_{2}$
  ($\it{A}=\mathrm{K},\mathrm{Tl},\mathrm{Rb}$)}.
\newblock \emph{\bibinfo{journal}{New J. Phys.}} \textbf{\bibinfo{volume}{15}},
  \bibinfo{pages}{093045} (\bibinfo{year}{2013}).
\newblock \urlprefix\url{http://dx.doi.org/10.1088/1367-2630/15/9/093045}.

\bibitem{Tsutsumi59}
\bibinfo{author}{Tsutsumi, K.}
\newblock \bibinfo{title}{The x-ray non-diagram lines ${K}\beta'$ of some
  compounds of the iron group}.
\newblock \emph{\bibinfo{journal}{J. Phys. Soc. Jpn.}}
  \textbf{\bibinfo{volume}{14}}, \bibinfo{pages}{1696} (\bibinfo{year}{1959}).
\newblock \urlprefix\url{http://journals.jps.jp/doi/abs/10.1143/JPSJ.14.1696}.

\bibitem{Vanko06}
\bibinfo{author}{Vank{\'o}, G.} \emph{et~al.}
\newblock \bibinfo{title}{Probing the $3d$ spin momentum with x-ray emission
  spectroscopy: the case of molecular-spin transitions}.
\newblock \emph{\bibinfo{journal}{Phys. Chem. B}}
  \textbf{\bibinfo{volume}{110}}, \bibinfo{pages}{11647}
  (\bibinfo{year}{2006}).
\newblock \urlprefix\url{http://dx.doi.org/10.1021/jp0615961}.

\bibitem{JMChen11}
\bibinfo{author}{Chen, J.~M.} \emph{et~al.}
\newblock \bibinfo{title}{Pressure dependence of the electronic structure and
  spin state in $\mathrm{Fe}_{1.01}\mathrm{Se}$ superconductors probed by x-ray
  absorption and x-ray emission spectroscopy}.
\newblock \emph{\bibinfo{journal}{Phys. Rev. B}} \textbf{\bibinfo{volume}{84}},
  \bibinfo{pages}{125117} (\bibinfo{year}{2011}).
\newblock \urlprefix\url{http://link.aps.org/doi/10.1103/PhysRevB.84.125117}.

\bibitem{Gretarsson11}
\bibinfo{author}{Gretarsson, H.} \emph{et~al.}
\newblock \bibinfo{title}{Revealing the dual nature of magnetism in iron
  pnictides and iron chalcogenides using x-ray emission spectroscopy}.
\newblock \emph{\bibinfo{journal}{Phys. Rev. B}} \textbf{\bibinfo{volume}{84}},
  \bibinfo{pages}{100509} (\bibinfo{year}{2011}).
\newblock \urlprefix\url{http://dx.doi.org/10.1103/PhysRevB.84.100509}.

\bibitem{Gretarsson13}
\bibinfo{author}{Gretarsson, H.} \emph{et~al.}
\newblock \bibinfo{title}{Spin-state transition in the fe pnictides}.
\newblock \emph{\bibinfo{journal}{Phys. Rev. Lett.}}
  \textbf{\bibinfo{volume}{110}}, \bibinfo{pages}{047003}
  (\bibinfo{year}{2013}).
\newblock
  \urlprefix\url{http://link.aps.org/doi/10.1103/PhysRevLett.110.047003}.

\bibitem{Hamalainen91}
\bibinfo{author}{H\"am\"al\"ainen, K.}, \bibinfo{author}{Siddons, D.~P.},
  \bibinfo{author}{Hastings, J.~B.} \& \bibinfo{author}{E.Berman, L.}
\newblock \bibinfo{title}{Elimination of the inner-shell lifetime broadening in
  x-ray-absorption spectroscopy}.
\newblock \emph{\bibinfo{journal}{Phys. Rev. Lett.}}
  \textbf{\bibinfo{volume}{67}}, \bibinfo{pages}{2850} (\bibinfo{year}{1991}).
\newblock \urlprefix\url{http://link.aps.org/doi/10.1103/PhysRevLett.67.2850}.

\bibitem{Dallera02}
\bibinfo{author}{Dallera, C.} \emph{et~al.}
\newblock \bibinfo{title}{New spectroscopy solves an old puzzle: the kondo
  scale in heavy fermions}.
\newblock \emph{\bibinfo{journal}{Phys. Rev. Lett.}}
  \textbf{\bibinfo{volume}{88}}, \bibinfo{pages}{196403}
  (\bibinfo{year}{2002}).
\newblock
  \urlprefix\url{http://link.aps.org/doi/10.1103/PhysRevLett.88.196403}.

\bibitem{Yamaoka14}
\bibinfo{author}{Yamaoka, H.} \emph{et~al.}
\newblock \bibinfo{title}{Role of valence fluctuations in the superconductivity
  of $\mathrm{Ce}$122 compounds}.
\newblock \emph{\bibinfo{journal}{Phys. Rev. Lett.}}
  \textbf{\bibinfo{volume}{113}}, \bibinfo{pages}{086403}
  (\bibinfo{year}{2014}).
\newblock
  \urlprefix\url{http://link.aps.org/doi/10.1103/PhysRevLett.113.086403}.

\bibitem{Fujita15}
\bibinfo{author}{Fujita, H.} \emph{et~al.}
\newblock \bibinfo{title}{Pressure dependence of superconductive transition
  temperature on $\mathrm{K}_{x}\mathrm{Fe}_{2-y}\mathrm{Se}_{2}$}.
\newblock \emph{\bibinfo{journal}{J. Phys.: Conf. Ser.}}
  \textbf{\bibinfo{volume}{592}}, \bibinfo{pages}{012070}
  (\bibinfo{year}{2015}).
\newblock \urlprefix\url{http://dx.doi.org/10.1088/1742-6596/592/1/012070}.

\bibitem{Tanaka15}
\bibinfo{author}{Tanaka, M.} \emph{et~al.}
\newblock \bibinfo{title}{Origin of the Higher-${T}_{\mathrm{c}}$ Phase in the
  $\mathrm{K}_{x}\mathrm{Fe}_{2-y}\mathrm{Se}_{2}$ system}.
\newblock \emph{\bibinfo{journal}{J. Phys. Soc. Jpn.}}
  \textbf{\bibinfo{volume}{85}}, \bibinfo{pages}{044710}
  (\bibinfo{year}{2016}).
\newblock \urlprefix\url{http://dx.doi.org/10.7566/JPSJ.85.044710}.

\bibitem{Bendele13}
\bibinfo{author}{Bendele, M.} \emph{et~al.}
\newblock \bibinfo{title}{Interplay of electronic and lattice degrees of
  freedom in ${A}_{1-x}\mathrm{Fe}_{2-y}\mathrm{Se}_{2}$ superconductors under
  pressure}.
\newblock \emph{\bibinfo{journal}{Phys. Rev. B}} \textbf{\bibinfo{volume}{88}},
  \bibinfo{pages}{180506} (\bibinfo{year}{2013}).
\newblock \urlprefix\url{http://dx.doi.org/10.1103/PhysRevB.88.180506}.

\bibitem{Yu14}
\bibinfo{author}{Yu, Z.} \emph{et~al.}
\newblock \bibinfo{title}{Conventional empirical law reverses in the phase
  transitions of 122-type iron-based superconductors}.
\newblock \emph{\bibinfo{journal}{Sci. Rep.}} \textbf{\bibinfo{volume}{4}},
  \bibinfo{pages}{7172} (\bibinfo{year}{2014}).
\newblock \urlprefix\url{http://dx.doi.org/10.1038/srep07172}.

\bibitem{Groot09}
\bibinfo{author}{de~Groot, F.}, \bibinfo{author}{Vanko, G.} \&
  \bibinfo{author}{Glatzel, P.}
\newblock \bibinfo{title}{The 1s x-ray absorption pre-edge structures in
  transition metal oxides}.
\newblock \emph{\bibinfo{journal}{Journal of Physics: Condensed Matter}}
  \textbf{\bibinfo{volume}{21}}, \bibinfo{pages}{104207}
  (\bibinfo{year}{2009}).
\newblock \urlprefix\url{http://stacks.iop.org/0953-8984/21/i=10/a=104207}.

\bibitem{Simonelli12}
\bibinfo{author}{Simonelli, L.} \emph{et~al.}
\newblock \bibinfo{title}{Coexistence of different electronic phases in the
  $\mathrm{K}_{0.8}\mathrm{Fe}_{1.6}\mathrm{Se}_{2}$ superconductor: A
  bulk-sensitive hard x-ray spectroscopy study}.
\newblock \emph{\bibinfo{journal}{Phys. Rev. B}} \textbf{\bibinfo{volume}{85}},
  \bibinfo{pages}{224510} (\bibinfo{year}{2012}).
\newblock \urlprefix\url{http://link.aps.org/doi/10.1103/PhysRevB.85.224510}.

\bibitem{Ying15}
\bibinfo{author}{Ying, J.-J.} \emph{et~al.}
\newblock \bibinfo{title}{Tripling the critical temperature of
  $\mathrm{K}\mathrm{Fe}_{2}\mathrm{As}_{2}$ by carrier switch}.
\newblock \emph{\bibinfo{journal}{arXiv:1501.00330}}  (\bibinfo{year}{2015}).
\newblock \urlprefix\url{http://arxiv.org/abs/1501.00330}.

\bibitem{Nakajima15}
\bibinfo{author}{Nakajima, Y.} \emph{et~al.}
\newblock \bibinfo{title}{High-temperature superconductivity stabilized by
  electron-hole interband coupling in collapsed tetragonal phase of
  $\mathrm{K}\mathrm{Fe}_{2}\mathrm{As}_{2}$ under high pressure}.
\newblock \emph{\bibinfo{journal}{Phys. Rev. B}} \textbf{\bibinfo{volume}{91}},
  \bibinfo{pages}{060508} (\bibinfo{year}{2015}).
\newblock \urlprefix\url{http://link.aps.org/doi/10.1103/PhysRevB.91.060508}.

\bibitem{Guterding15}
\bibinfo{author}{Guterding, D.}, \bibinfo{author}{Backes, S.},
  \bibinfo{author}{Jeschke, H.~O.} \& \bibinfo{author}{Valent\'{\i}, R.}
\newblock \bibinfo{title}{Origin of the superconducting state in the collapsed
  tetragonal phase of $\mathrm{KFe}_{2}\mathrm{As}_{2}$}.
\newblock \emph{\bibinfo{journal}{Phys. Rev. B}} \textbf{\bibinfo{volume}{91}},
  \bibinfo{pages}{140503} (\bibinfo{year}{2015}).
\newblock \urlprefix\url{http://dx.doi.org/10.1103/PhysRevB.91.140503}.

\bibitem{Ozaki13}
\bibinfo{author}{Ozaki, T.} \emph{et~al.}
\newblock \bibinfo{title}{Evolution of superconductivity in isovalent
  {Te}-substituted $\mathrm{K}\mathrm{Fe}_{2-y}\mathrm{Se}_{2}$ crystals}.
\newblock \emph{\bibinfo{journal}{Supercond. Sci. Technol.}}
  \textbf{\bibinfo{volume}{26}}, \bibinfo{pages}{055002}
  (\bibinfo{year}{2013}).
\newblock \urlprefix\url{http://stacks.iop.org/0953-2048/26/i=5/a=055002}.

\bibitem{Mao76}
\bibinfo{author}{Mao, H.-K.} \& \bibinfo{author}{Bell, P.~M.}
\newblock \bibinfo{title}{High-pressure physics: The 1-megabar mark on the ruby
  ${R}_{1}$ static pressure scale}.
\newblock \emph{\bibinfo{journal}{Science}} \textbf{\bibinfo{volume}{191}},
  \bibinfo{pages}{851} (\bibinfo{year}{1976}).
\newblock \urlprefix\url{http://dx.doi.org/10.1126/science.191.4229.851}.

\bibitem{Hammersley96}
\bibinfo{author}{Hammersley, A.~P.}, \bibinfo{author}{Svensson, S.~O.},
  \bibinfo{author}{Hanfland, M.}, \bibinfo{author}{Fitch, A.~N.} \&
  \bibinfo{author}{Hausermann, D.}
\newblock \bibinfo{title}{Two-dimensional detector software: From real detector
  to idealised image or two-theta scan}.
\newblock \emph{\bibinfo{journal}{High Pressure Research}}
  \textbf{\bibinfo{volume}{14}}, \bibinfo{pages}{235} (\bibinfo{year}{1996}).
\newblock \urlprefix\url{http://dx.doi.org/10.1080/08957959608201408}.

\end{thebibliography}

\section*{Acknowledgements}
The experiments were performed at Taiwan beam line BL12XU and BL12B2 at SPring-8 under SPring-8 Proposals No.~2013B4127, No.~2013B4156, {\&} 2014A4257 (corresponding NSRRC Proposal No. 2013-3-007). We are grateful to Yumiko Zekko, Satomi Kawase and Yu Ohta in Kwansei Gakuin University and Yuki Sumi in Doshisha University for their help in the experiment. We deeply thank Young-June Kim at University of Toronto for the preparation of the FeCrAs  sample. We also deeply appreciate Jin-Ming Chen and Jenn-Min Lee for the use of the diamond anvil cell system in the XRD experiment and Akihiko Machida for use of the glove box in JAERI. This work was partly supported by JSPS KAKENHI Grant Number 15K05194 and 26400322. This work at UT Austin was supported as part of EFree, an Energy Frontier Research Center funded by the U. S. Department of Energy Office of Science, Office of Basic Energy Sciences under Award DE-SC0001057.

\end{document}

\title{Supplemental information: Origin of Pressure-induced Superconducting Phase in $\mathrm{K}_{x}\mathrm{Fe}_{2-y}\mathrm{Se}_{2}$ studied by Synchrotron X-ray Diffraction and Spectroscopy}

\author{Yoshiya Yamamoto} 
  \affiliation{Graduate School of Science and Technology, Kwansei Gakuin University, 2-1 Gakuen, Sanda, Hyogo 669-1337, Japan}
\author{Hitoshi Yamaoka}
  \affiliation{RIKEN SPring-8 Center, RIKEN, 1-1-1 Kouto, Mikazuki, Sayo, Hyogo 679-5148, Japan}
\author{Masashi Tanaka}
  \affiliation{MANA, National Institute for Materials Science, 1-2-1 Sengen, Tsukuba, Ibaraki 305-0047, Japan}
\author{Hiroyuki Okazaki}
  \affiliation{MANA, National Institute for Materials Science, 1-2-1 Sengen, Tsukuba, Ibaraki 305-0047, Japan}
  \affiliation{Advanced Institute for Materials Research, Tohoku University, 2-1-1 Katahira, Aoba, Sendai, Miyagi 980-8577, Japan}
\author{Toshinori Ozaki}
  \affiliation{MANA, National Institute for Materials Science, 1-2-1 Sengen, Tsukuba, Ibaraki 305-0047, Japan}
  \affiliation{Graduate School of Science and Technology, Kwansei Gakuin University, 2-1 Gakuen, Sanda, Hyogo 669-1337, Japan}
\author{Yoshihiko Takano}
  \affiliation{MANA, National Institute for Materials Science, 1-2-1 Sengen, Tsukuba, Ibaraki 305-0047, Japan}
\author{Jung-Fu Lin}
  \affiliation{Department of Geological Sciences, The University of Texas at Austin, Austin, Texas 78712, USA}
  \affiliation{Center for High Pressure Science and Technology Advanced Research (HPSTAR), Shanghai 201203, China}
\author{Hidenori Fujita}
  \affiliation{Center for Science and Technology under Extreme Conditions (KYOKUGEN), 
Graduate School of Engineering Science, Osaka University, Toyonaka, Osaka 560-8531, Japan}
\author{Tomoko Kagayama}
  \affiliation{Center for Science and Technology under Extreme Conditions (KYOKUGEN), 
Graduate School of Engineering Science, Osaka University, Toyonaka, Osaka 560-8531, Japan}
\author{Katsuya Shimizu}
  \affiliation{Center for Science and Technology under Extreme Conditions (KYOKUGEN), 
Graduate School of Engineering Science, Osaka University, Toyonaka, Osaka 560-8531, Japan}
\author{Nozomu Hiraoka}
  \affiliation{National Synchrotron Radiation Research Center, Hsinchu 30076, Taiwan}
\author{Hirofumi Ishii}
  \affiliation{National Synchrotron Radiation Research Center, Hsinchu 30076, Taiwan}
\author{Yen-Fa Liao}
  \affiliation{National Synchrotron Radiation Research Center, Hsinchu 30076, Taiwan}
\author{Ku-Ding Tsuei}
  \affiliation{National Synchrotron Radiation Research Center, Hsinchu 30076, Taiwan}
\author{Jun'ichiro Mizuki}
  \affiliation{Graduate School of Science and Technology, Kwansei Gakuin University, 2-1 Gakuen, Sanda, Hyogo 669-1337, Japan}
 
\maketitle

Here we show additional information about the results of the XRD and the PFY-XAS and a correlation of the magnetic moment with $T_{\mathrm{c}}$. 

\section{X-ray diffraction}  
\begin{figure}[h]
  \includegraphics{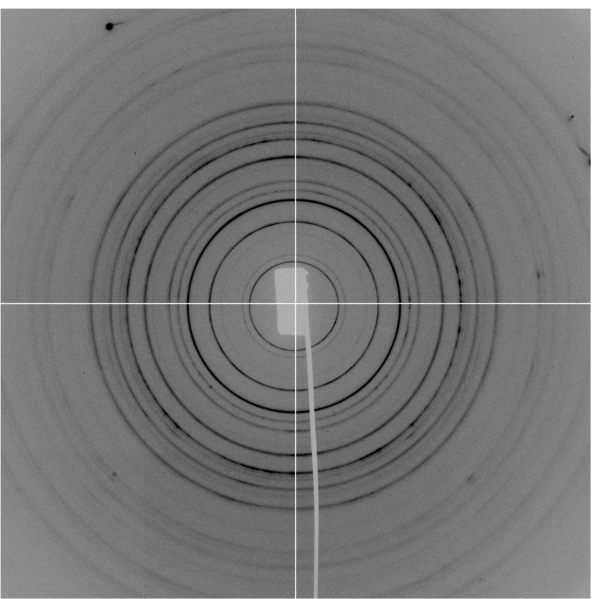}
  \caption{(Color online) An example of the 2D XRD pattern of the quenched sample.}
  \label{fig:XRDimage}
\end{figure}

Figure~\ref{fig:XRDimage} shows an example of the 2D XRD pattern.
There is no spot peak with preferred orientation. 
Note that this image includes the strong peak from diamond.
We integrated the image with masking by using FIT2D program.\cite{Hammersley96}

\begin{figure}[h]
  \includegraphics{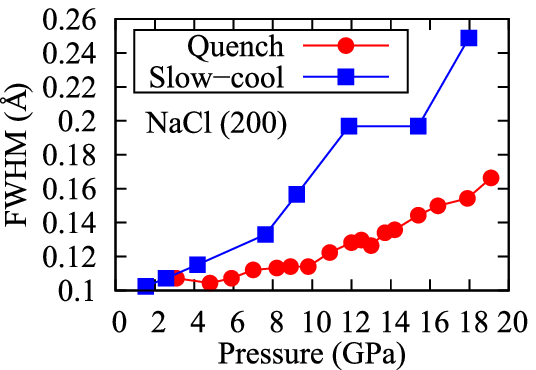}
  \caption{(Color online) Full width at half maximum (FWHM) of the (200) peak of NaCl used as the pressure medium of the diamond anvil cell. Red closed circle and blue closed square correspond to the quenched and slow-cooled samples, respectively.}
  \label{fig:NaCl_FWHM}
\end{figure}

Figure~\ref{fig:NaCl_FWHM} shows full width at half maximum (FWHM) of the (200) peak of NaCl used as the pressure medium of the diamond anvil cell.
Change in the FWHM of (200) peak increases smoothly with pressure and no sudden change is observed in the pressure range measured. 
Therefore, the trend of the pressure-induced change in the lattice constant along the $c$-axis around 12 GPa is reliable. 

\begin{figure}
  \includegraphics{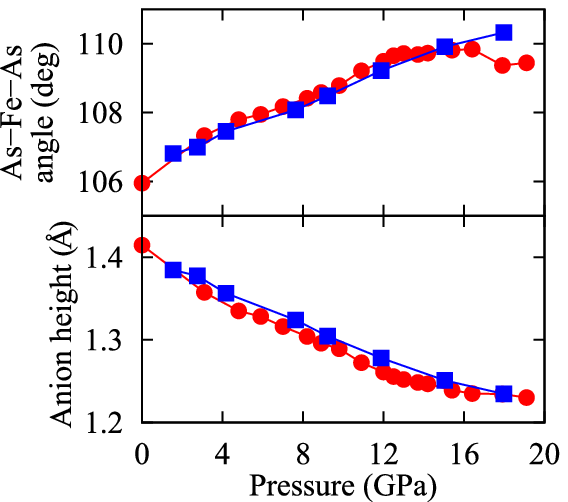}
  \caption{(Color online) Anion height and As-Fe-As bond angle calculated from lattice constants of the quenched samples  (closed square) and slow-cooled sample (closed circle).}
  \label{fig:angle&hight}
\end{figure}

Figure~\ref{fig:angle&hight} shows the anion height and the As-Fe-As bond angle estimated from the lattice constants.
In $\mathrm{EuFe}_{2}\mathrm{As}_{2}$, Wyckoff positions change with pressure, especially $z$ position changes dynamically.\cite{Yu14}
However, roughly, the bond angle takes an optimum value of $\sim$109$^{\circ}$ at the SC II phase and the anion height does an optimum value of $\sim$1.38 {\AA} at the SC I phase.\cite{Mizuguchi10} 
The bond angle increases with pressure, while the anion height decreases monotonically. 
Both the trend of the changes in the bond angle and anion height seems to be gentle above 12 GPa. 

\section{PFY-XAS}  
\begin{figure*}[h]
  \includegraphics{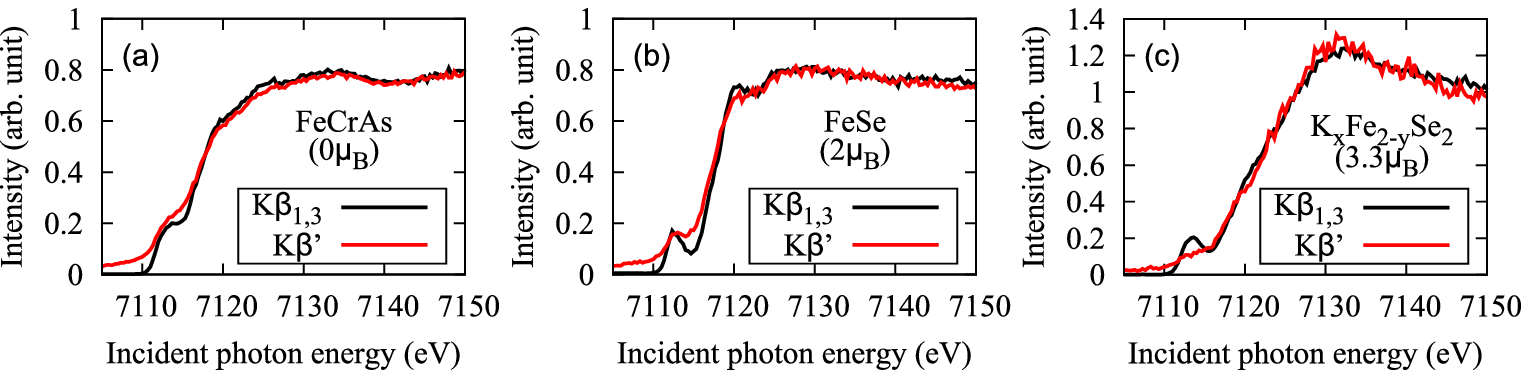}
  \caption{(Color online) PFY-XAS spectra of (a) FeCrAs ($0\mu_{\mathrm{B}}$) and (b) FeSe ($2\mu_{\mathrm{B}}$) and (c) $\mathrm{K}_{x}\mathrm{Fe}_{2-y}\mathrm{Se}_{2}$ ($3.3\mu_{\mathrm{B}}$) set the emitted photon energy to the $K\beta_{1,3}$ peak (red line) and $K\beta'$ peak (blue line).}
  \label{fig:PFY-XAS_Kb-dep}
\end{figure*}

Figure~\ref{fig:PFY-XAS_Kb-dep} shows PFY-XAS spectra of FeCrAs ($0\mu_{\mathrm{B}}$), FeSe ($2\mu_{\mathrm{B}}$) and $\mathrm{K}_{x}\mathrm{Fe}_{2-y}\mathrm{Se}_{2}$ ($3.3\mu_{\mathrm{B}}$).
We measured two kinds of the PFY-XAS spectra by setting the emitted photon energies to the $K\beta_{1,3}$ and $K\beta'$ peaks, respectively. 
The intensity of the pre-edge peak of the PFY-XAS spectra at the $K\beta'$ peak (red lines in Fig. 3) decreases with increasing magnetic moment.
Both the PFY-XAS spectra show same feature above the $K$-absorption energy. 
It is considered that the difference of the feature of the pre-edge peak reflects the spin state. 
The intensity of the pre-edge peak of the spectra at the $K\beta'$ peak is weak normally in the high-spin state.\cite{Yamaoka04} 
Figure 4 shows higher-magnetic moment correlates to the lower intensity of the pre-edge peak of the PFY-XAS spectra at the $K\beta'$ peak, reasonably corresponding to the higher-spin state. 

The pre-edge peak intensity at the $K\beta_{1,3}$ increases with pressure, corresponding to the change in the spin state from the high-spin to low-spin state.
This also suggests that Fe $3d$-Se $4p$ hybridization becomes stronger with pressure and thus the DOS near the Fermi surface is higher at the high pressure phase.

\begin{figure*}[h]
  \includegraphics{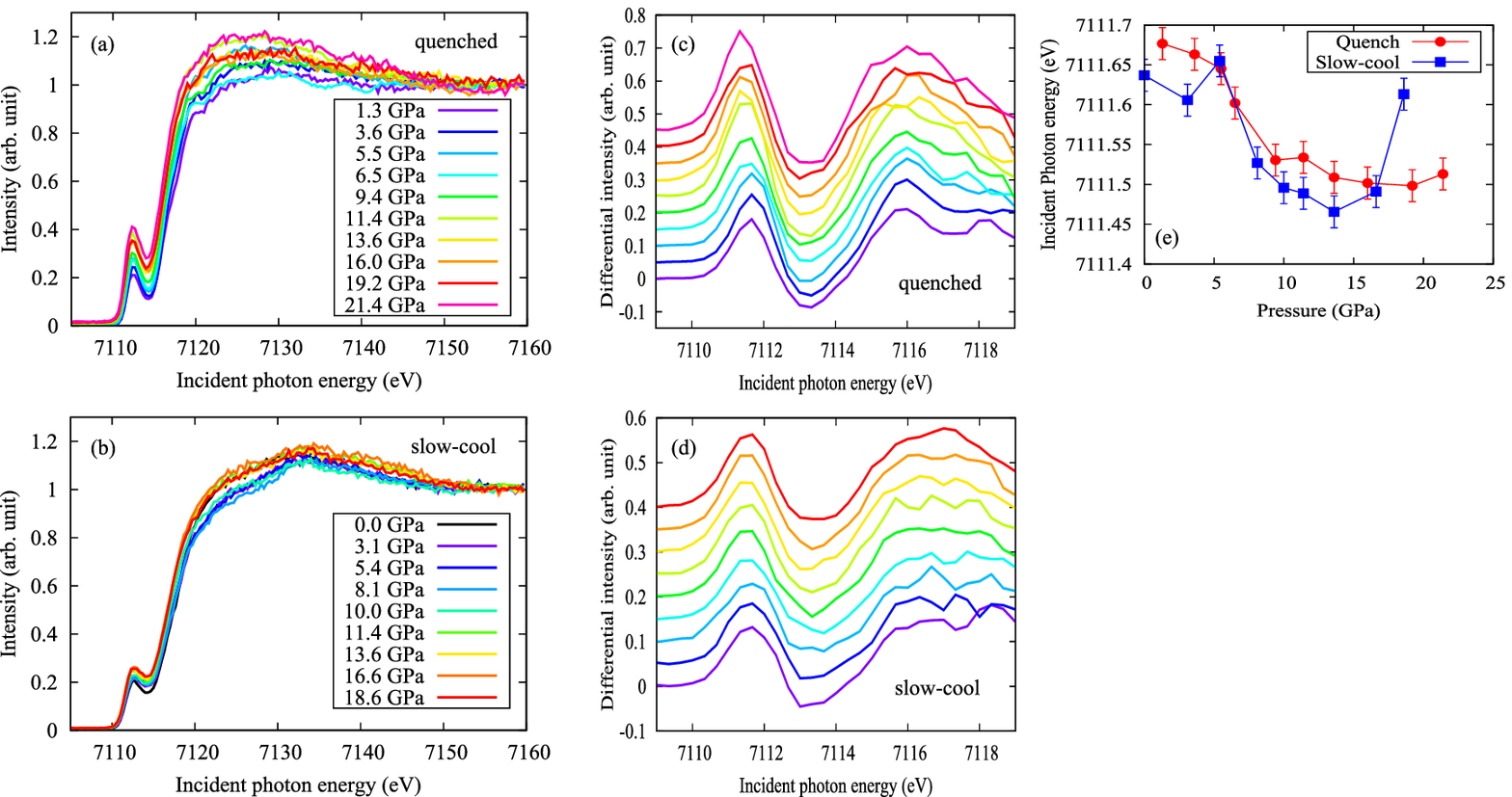}
  \caption{(Color online) Pressure evolution of the PFY-XAS spectra of (a) the quenched and (b) the slow-cooled samples. In both the quenched and slow-cooled samples, the pre-edge peak intensity increases with pressure and the edge position moves towered the low energy. (c) and (d) Pressure evolution of the differential of the PFY-XAS spectra around the pre-edge peak. Upper and lower spectra correspond to high and low pressures, respectively. (e) Pressure evolution of the peak position of the differential PFY-XAS spectra around 7111.5 eV. }
  \label{fig:chemicalpotential}
\end{figure*}

Figure~\ref{fig:chemicalpotential} shows the pressure dependence of the PFY-XAS spectra.
In Figs.~\ref{fig:chemicalpotential}(c) and \ref{fig:chemicalpotential}(d) we show the pressure dependence of the differential of the PFY-XAS spectra near the pre-edge peak region. 
The pressure-induced change in the peak position of the differential of the PFY-XAS spectra indicates that the inflection point changes to lower energy with pressure as shown in Fig.~\ref{fig:chemicalpotential}(e).
The evolution of the inflection point corresponds to that of the chemical potential.\cite{Bendele13}
Thus the chemical potential decreases with pressure at SC II phase.
Bendele {\it et al.} showed that the chemical potential at the SC II phase was lower than that at the SC I phase, suggesting the increase of the density of states with pressure.\cite{Bendele13}
Our results show a similar trend except 7-11 GPa, where a gradual change in the chemical potential is suggested. 
The peak positions gradually drop with pressure above 5 GPa at the SC I phase and do not change much at the SC II phase above 10 GPa.

\begin{figure}[h]
  \includegraphics{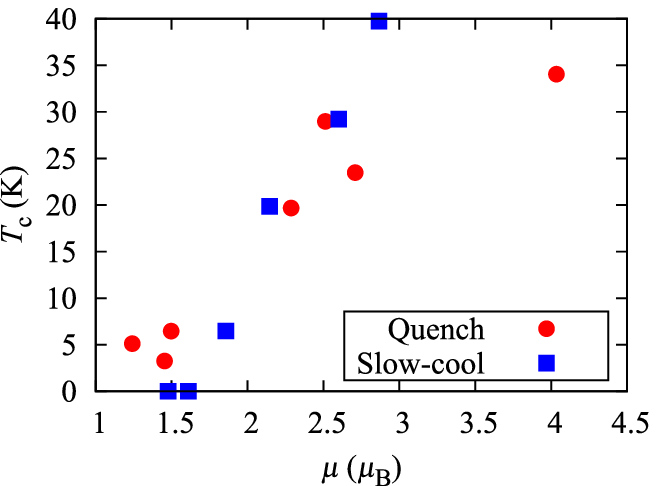}
  \caption{(Color online) Correlation between $\mu$ and $T_{\mathrm{c}}$} at the SC I phase. 
  \label{fig:mu-Tc}
\end{figure}

Figure~\ref{fig:mu-Tc} shows a correlation of the magnetic moment $\mu$ and $T_{\mathrm{c}}$ at the SC I phase.
The data of $T_{\mathrm{c}}$ in the $p$-$T$ phase diagram was fitted  and $T_{\mathrm{c}}$ at a given pressure was estimated. 
There is an apparent correlation between the magnetic moment and $T_{\mathrm{c}}$. 
Theoretically, it was suggested that a spin-lattice coupling possibly occurs in the Fe pnictide superconductors and the Fe local moment correlates to the Fe-As layer separation, i. e. the anion height.\cite{Egami10} 
The anion height as well as $T_{\mathrm{c}}$ decreases with pressure at the SC I phase as shown in Fig.~\ref{fig:angle&hight}. 
Thus, the above theory can explain that the pressure-induced change in the local magnetic moment derived from the $K\beta$ XES spectra reasonably correlates to the superconductivity at the SC I phase.  
But further theoretical and experimental studies will be required to understand the correlation between the superconductivity and local magnetism.

%\bibliography{Suppl_KxFe2-ySe2}